# The citation advantage of foreign language references for Chinese social science papers

Kaile Gong[1,2], Juan Xie[1], Ying Cheng[1], Vincent Larivière[3], Cassidy R. Sugimoto[2,*]

[1] School of Information Management, Nanjing University, 163 Xianlin Road, Nanjing, Jiangsu 210023, China
[2] School of Informatics, Computing, and Engineering, Indiana University Bloomington, 901 East 10th Street, Bloomington, IN 47408, USA
[3] École de bibliothéconomie et des sciences de l'information, Université de Montréal, C.P. 6128, Succ. Centre-Ville, Montréal, QC H3C 3J7, Canada

[*] Corresponding author. Email: sugimoto@indiana.edu

**Abstract** Contemporary scientific exchanges are international, yet language continues to be a persistent barrier to scientific communication, particularly for non-native English-speaking scholars. Since the ability to absorb knowledge has a strong impact on how researchers create new scientific knowledge, a comprehensive access to and understanding of both domestic and international scientific publications is essential for scientific performance. This study explores the effect of absorbed knowledge on research impact by analyzing the relationship between the language diversity of cited references and the number of citations received by the citing paper. Chinese social sciences are taken as the research object, and the data, 950,302 papers published between 1998 and 2013 with 8,151,327 cited references, were collected from the Chinese Social Sciences Citation Index (CSSCI). Results show that there is a stark increase in the consumption of foreign language material within the Chinese social science community, and English material accounts for the vast majority of this consumption. Papers with foreign language references receive significantly more citations than those without, and the citation advantage of these internationalized work holds when we control for characteristics of the citing papers, such as the discipline, prestige of journal, prestige of institution, and scientific collaboration. However, the citation advantage has decreased from 1998 to 2008, largely as an artifact of the increased number of papers citing foreign language material. After 2008, however, the decline of the citation advantage subsided and became relatively stable, which suggests that incorporating foreign language literature continues to increase scientific impact, even as the scientific community itself becomes increasingly international. However, internationalization is not without concerns: the work closes with a discussion of the benefits and potential problems of the lack of linguistic diversity in scientific communication.

# 1. Introduction

Contemporary scientific exchanges are international. Researchers collaborate with colleagues from other countries, migrate for short- or long-term positions, and attend academic conferences in far flung places. These international exchanges have a positive influence on scientific impact (Aksnes 2003; Sugimoto et al. 2017). However, language barriers remain one of the main obstacles to international scientific communication and dissemination of knowledge, despite increasingly efficient translation tools. This barrier is growing for non-native English-speaking scholars, as scientists increasingly publish in English, even in the social sciences and humanities (Kulczycki et al. 2017; Larivière 2018). This language barrier is one of both access and understanding: scholars are only able to access the information in languages in which they have high literacy and, even with advanced literacy, full absorption of content may be lessened for non-native speakers (Liu 2017). An old Chinese saying states: "*Only by learning extensively and accumulating profound knowledge can one be ready to achieve something*". Taken as a metaphor for scientific success, one might hypothesize that without full access to the scientific literature, success may be limited. In short, the lack of ability to absorb international scientific literature may lessen the potential for scientific impact.

Citations are a generally accepted indicator of scientific impact; they indicate the degree to which a cited article influenced subsequent scientific work (Bornmann and Daniel 2008). However, they are often misconstrued as indicators of quality (Borgman and Furner 2002), despite demonstrations to the contrary (Nieminen et al. 2006). This is largely due to the wide heterogeneity in referencing motivation (Bornmann 2016; Wang et al. 2018), wherein a work may be cited not due to the quality of the work but for a variety of factors. The main underlying factor was evoked by Merton's (1988) characterization of citations as "pellets of peer recognition"; that is, a reference serves the function of indicators of the impact (whether positive or negative) of the past work upon the present. In this way, references can be seen as a proxy for knowledge transmission and absorption. References and citations are two sides of the same coin: if references are made for a variety of reasons, so too are citations received. In a comprehensive review, Tahamtan et al. (2016) identified several factors that affect papers' citation counts, including paper-, journal-, and author-related factors.

One paper-related factor is the references within a paper. Several studies have explored the relationship between a paper's citation count and the characteristics of its references. Authors have demonstrated a positive relationship between citation count and number of references (e.g., Antoniou et al. 2015; Biscaro and Giupponi 2014; Bornmann et al. 2014; Chen 2012; Didegah and Thelwall 2013; Falagas et al. 2013; Gargouri et al. 2010; Haslam and Koval 2010; Lokker et al. 2008; Mou et al. 2018; Onodera and Yoshikane 2015; Roth et al. 2012; So et al. 2015; van Wesel et al. 2014; Yu and Yu 2014; Yu et al. 2014), citedness of references (Bornmann et al. 2012; Mou et al. 2018), age of references (Roth et al. 2012), interdisciplinarity of references (Chakraborty et al. 2014), and document types of references (Miranda and Garcia-Carpintero 2018; Mou et al. 2018). These previous studies suggest that the number, diversity, prestige, and timeliness of absorbed knowledge can affect scientific impact of a published work.

Discovering new scientific knowledge combines both understanding and absorbing the current state of a field, but also creating, inventing, and obtaining new results, concepts, and ideas (Cheng 2009; Qiu et al. 2016). Therefore, the ability to absorb current knowledge has a significant impact on how researchers create new knowledge (Cohen and Levinthal 1990). It may be hypothesized that language barriers—evidenced by the diversity in the language of cited references—may lessen the potential for scientific impact. Previous research has found several factors that influence papers citing foreign language literature (Gong et al. 2018). For example, in terms of publications written in Chinese, there are more foreign language literature cited by papers published in prestigious journals (Gong et al. 2018). Since papers published in prestigious journals tend to receive more citations (Callaham et al. 2002), one can hypothesize that this effect is also observed at the level of Chinese literature. In addition, other factors such as the discipline, prestige of institution and scientific collaboration were also widely proved to influence both citing behavior and citations received (Bornmann and Daniel 2008; Tahamtan et al. 2016). To better understand the factors affecting citation rates, this paper aims to explore the effect of absorbed knowledge on research impact by analyzing the relationship between the language diversity of cited references and the number of citations received by the citing paper. Specifically, three research questions are addressed in the current study:

RQ1. Do papers with foreign language references receive more citations than those without?

RQ2. Does the citation advantage of papers with foreign language references decrease as their proportion increases?

RQ3. Does the citation advantage of papers with foreign language references remain after controlling the widely-recognized characteristics of citing papers?

## 2. Data and Methods

### 2.1 Research objects

Chinese social sciences are selected for the analysis for three reasons. Firstly, compared with natural scientists who have universal research objects and paradigms, social scientists tend to publish papers in local journals in their native language because their research purposes, problems, and objects are often closely related to the social environment and culture where they reside, and tend to have greater relevance within that specific region (Gingras and Mosbah-Natanson 2010; Warren 2004). Correspondingly, the knowledge on which they rely is more likely to be in their native language (Kulczycki et al. 2017; Larivière 2018; Yitzhaki 1998). Therefore, the issue of language may be more applicable to the social, rather than the natural sciences.

Secondly, over the course of the 20$^{th}$ century, English became the scientific lingua franca (Gordin 2015; Montgomery 2013). However, the 21$^{st}$ century is witnessing a strong increasing in the scientific output of China. China has been the largest producer of scholarly publications in the world since 2016 (National Science Foundation 2018), as measured by Scopus, which has a demonstrated bias for English-language material. The fact that a non-English-speaking country is the largest producer of scholarly papers, combined with the high homophily of referencing practices observed across country (Larivière et al. 2019)—bears examination of how the referencing practices of Chinese papers have evolved over time, particularly with regards to referencing English-language material.

Thirdly, China is increasingly active in social science research—despite not being as central to international scientific exchanges as in the natural and medical sciences (Liu et al. 2015; Zhou et al. 2009). Such growth can be associated with increased international interest in China, as well as, reciprocally, of the increasing openness of Chinese scholars in international academic exchange. While we know the extent to which Chinese social sciences researchers publishing in international journals—as represented by the Web of Science and Scopus—little is known on the use of English-language journals in Chinese literature of the social sciences.

## 2.2 Data collection

The data were collected in April 2019 from the Chinese Social Sciences Citation Index (CSSCI), which was created in 1998 by the Institute for Chinese Social Science Research and Assessment of Nanjing University. CSSCI was chosen as the data source because it is widely recognized as a China's leading, authoritative, and comprehensive database for scholarly citations, and it includes more than 600 high-impact journals in the fields of the humanities and social sciences (Wang et al. 2016). These journals are selected from among more than 2700 Chinese language journals in the field of humanities and social sciences based on quantitative indicators and qualitative peer review (Institute for Chinese Social Science Research and Assessment of Nanjing University 2016). The CSSCI database records rich metadata, such as title, authors, institutions, discipline, journal, keywords, references, and citation count within the CSSCI, which allows for the compilation of bibliometrics indicators.

CSSCI-indexed journals are sponsored by Chinese institutions and primarily published in China by authors affiliated by Chinese institutions. By 2018, less than 2% (n= 25,504; 1.49%) of CSSCI-indexed papers have authors from other countries and only 454 papers are not written in Chinese, accounting for 0.027% of the total papers. Among these 454 papers, only 38 have authors from other countries. Therefore, this strongly represents a nationalistic output of scholarly publications.

A total of 950,302 papers (and 8,151,327 cited references) from all 13 social science disciplines indexed by CSSCI from 1998 to 2013 are included, all of which are written in the Chinese language. In order to have a five-year citation window for all papers, papers published 2014 onwards are excluded from our analysis. Publications in this study are categorized into one or two of 13 disciplines: Economics (Econ), Education (Ed), Environmental Studies (Evs), Human & Economic Geography (Geo), Journalism & Communication (Jcom), Law (Law), Library & Information Science (Lis), Management (Mgmt), Political Science (Pols), Psychology (Psy), Sociology (Soc), Sports Science (Sps), and Statistics (Stat).

## 2.3 Papers' features coding

Each paper is coded for the following seven items:
1. ***With or without foreign language references***: Code the paper as "Papers with foreign language references" (PFLR) or "Papers without foreign language references" (PNLR). The identification of each reference's language is the basis of this study. The language of each cited reference is tagged as a number in CSSCI dataset (Su et al. 2014). The tagging rule is: Chinese-01, English-02, German-03, French-04, Russian-05, Japanese-06, Others-07, Translation-09 (to Chinese), and Korean-99. Using these data, all citing papers are coded into one of two binary categories according to the following rules: coding the citing paper as PFLR if it has at least one reference tagged as 02, 03, 04, 05, 06, 07 or 99; otherwise, coding it as PNLR. Hereby, 337,043 (35.5%) papers are coded as PFLR, and 613,259 (64.5%) papers are coded as PNLR. Meanwhile, Figure 1 shows that English references account for the vast majority (96.4%) of foreign language references, and the percentage of English references continued to raise while that of other languages decreased. Thus, to some extent, the foreign language references almost exclusively refer to English references in this study.

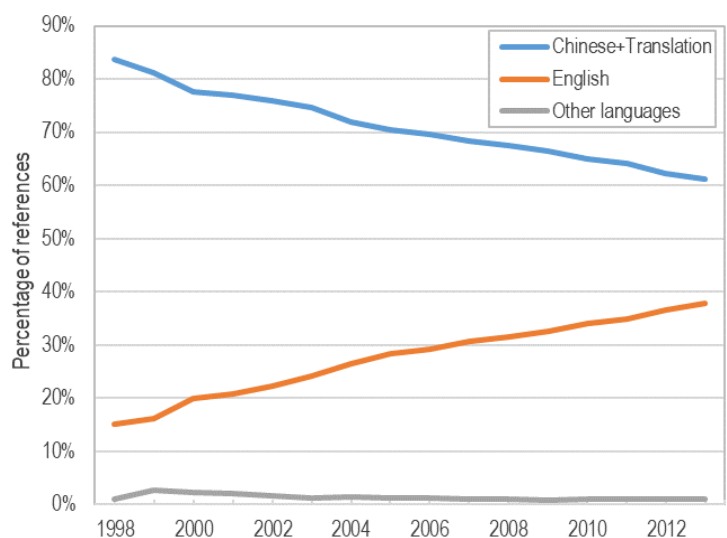

**Fig. 1** The percentage of references in different languages, by year

2. *Published year*: Code the paper by the published year. The number and percentage of papers in each year are shown in Table 1.

**Table 1** Number and percentage of papers, by published year

| Year | Number of papers | Percentage of papers |
|------|------------------|----------------------|
| 1998 | 42,095 | 4.4 |
| 1999 | 47,566 | 5.0 |
| 2000 | 44,781 | 4.7 |
| 2001 | 46,943 | 4.9 |
| 2002 | 49,974 | 5.3 |
| 2003 | 53,020 | 5.6 |
| 2004 | 57,804 | 6.1 |
| 2005 | 60,914 | 6.4 |
| 2006 | 65,815 | 6.9 |
| 2007 | 68,624 | 7.2 |
| 2008 | 71,864 | 7.6 |
| 2009 | 72,576 | 7.6 |
| 2010 | 68,101 | 7.2 |
| 2011 | 68,111 | 7.2 |
| 2012 | 66,864 | 7.0 |

| 2013 | 65,250 | 6.9 |
|---|---|---|

3. ***Discipline***: Code the paper as the disciplines that were tagged by CSSCI in the database. Since CSSCI tagged each paper into at most two disciplines, each paper in this study is associated with one or two discipline code(s). As noted above, there are 13 disciplines included in this study, the number and percentage of papers from each discipline are shown in Table 2. What needs illustration is that the sum of numbers is greater than 950,302 and the sum of percentages is greater than 100% is because of the multi-disciplinary coding rule for each paper. We use full counting in the analyses by discipline.

Table 2 Number and percentage of papers, by discipline

| Discipline | Number of papers | Percentage of papers |
|---|---|---|
| Econ | 318,284 | 33.5 |
| Ed | 132,449 | 13.9 |
| Mgmt | 122,141 | 12.9 |
| Lis | 82,307 | 8.7 |
| Pols | 82,062 | 8.6 |
| Jcom | 76,786 | 8.1 |
| Law | 75,158 | 7.9 |
| Soc | 38,600 | 4.1 |
| Sps | 30,180 | 3.2 |
| Evs | 21,091 | 2.2 |
| Psy | 18,347 | 1.9 |
| Geo | 5,412 | 0.6 |
| Stat | 5,208 | 0.5 |

4. ***Prestige of journal***: Code the paper as "Leading journals" or "Other journals" according to whether its journal belongs to the *Catalogue of Leading Journals in Humanities and Social Sciences*[1]. Hereby, 48,713 (5.1%) papers are coded as "Leading journals" and 901,589 (94.9%) papers are coded as "Other journals".

5. ***Prestige of institution***: Code the paper as "Top institutions" or "Other institutions" according to the authors' institution(s). If there is at least one institution listed among the 39 top universities in *Project 985*[2] or among the research institutions affiliated to two top research organizations from the Chinese Academy of Sciences[3] and Chinese Academy of

---

[1] The *Catalogue of Leading Journals in Humanities and Social Sciences* was compiled by the Social Sciences Department of Nanjing University. It contains the 31 most prestigious Chinese journals in the field of humanities and social sciences (Office of Humanities and Social Sciences of Nanjing University 2017). The catalogue was originally used by Nanjing University to evaluate the research performance of its faculty and students, but since there is no unified catalogue of top journals in China, it has gradually become one of the influential criteria for determining journal prestige in China (Chen 2016).

[2] *Project 985* is a project that was established by Chinese government in May 1998 to promote the development and reputation of the Chinese higher education system by founding world-class universities in the 21st century (Zong and Zhang 2017). There are 39 top Chinese universities sponsored by *Project 985* (Ministry of Education of the People's Republic of China 2006).

[3] The Chinese Academy of Sciences (CAS) is the world's largest research organization, comprising around 49,000 professional researchers working in more than 100 institutes (Yang et al. 2015), and has been consistently ranked

Social Sciences[4], in the institution(s) list, the paper is coded as "Top institutions". Otherwise, it's coded as "Other institutions". Hereby, 327,847 (34.5%) papers are coded as "Top institutions" and 622,455 (65.5%) papers are coded as "Other institutions".

6. ***Scientific collaboration***: Code the paper as "Co-authorship" if there are two or more authors in the byline; otherwise, code it as "Single authorship". Hereby, 371,167 (39.1%) papers are coded as "Co-authorship" and 579,135 (60.9%) papers are coded as "Single authorship".

7. ***Citation count***: The 2-year cumulative citation count (C2) and 5-year cumulative citation count (C5) within CSSCI are calculated as the basic citation indicators of each paper. Setting two different citation windows is partly because papers in different disciplines have various citation half-life, it may bias the results if only one citation window is used. In addition, the citation half-life of papers in social sciences is generally longer than natural sciences, it's better to set one more citation window that is longer than the commonly used - two years. Therefore, this study uses the citation window (2 and 5 years), respectively, to calculate the C2 and C5. The average value of C2 is 0.55 and that of C5 is 1.24.

## 2.4 Analysis

For the entire set of papers, we compile citations per paper (CPP), which is a common indicator used in citation analysis (see Eq. 1). This can be denoted:

$$\text{CPP} = \frac{C}{N} \quad (1)$$

where N is the total amount of papers in a certain set and C is the amount of citations received by these papers. The CPP therefore describes the arithmetic average of citation count of a certain paper set, and reflects the scholarly impact of these research outputs. Based on the C2 and C5, this study calculates the CPP with two citation windows, that is, 2-year cumulative citations per paper (CPP2) and 5-year cumulative citations per paper (CPP5).

Citation advantage (CA) represents the citation (dis)advantage of papers with foreign language references over those that do not contain such references. It is therefore used to measure the impact of language diversity of absorbed knowledge on research performance (see Eq. 2). The equation can be denoted as:

$$\text{CA} = \frac{\text{CPP}_{PFLR}}{\text{CPP}_{PNLR}} = \frac{C_{PFLR}/N_{PFLR}}{C_{PNLR}/N_{PNLR}} \quad (2)$$

where $\text{CPP}_{PFLR}$ is the average citation count of papers with foreign language references and

---

among the top research organizations around the world (Nature index 2018). Although it concentrates on the natural sciences, it has published more than 18,000 Chinese papers and 10,000 international papers in the field of social sciences (the statistical results are based on the CSSCI and SSCI).

[4] The Chinese Academy of Social Sciences (CASS) is the premier academic organization and comprehensive research center of China in the fields of philosophy and social sciences. It is affiliated with the PRC's State Council and consists of 31 research institutes and 45 research centers, which carry out research activities covering nearly 300 sub-disciplines (Du 2019).

CPP$_{PNLR}$ is the average citation count of papers without foreign language references. CA>1 means papers with foreign language references receive more citations than those without, on average, and suggests the language diversity of absorbed knowledge has a positive impact on research performance. CA<1 means the opposite. Based on the CPP2 and CPP5, this study calculates the CA with two citation windows. These are 2-year (CA2) and 5-year (CA5) cumulative citation advantage. As a comparative indicator, citation advantage (CA) reflects the quantitative advantage of one paper set's average citation count over another's, but there is no clear threshold to define the significance of this advantage. Therefore, further statistical tests are necessary to confirm if there is a significant difference in citation count between papers with foreign language references and those without. The Mann-Whitney U test is adopted in this study because the test variable, citation count, does not conform to normal distribution (Thelwall 2016) and the grouping variable "With or without foreign language references" is binary.

In order to answer RQ2, this study calculates the annual proportion of papers with foreign language references (%PFLR), the annual CA2 and the annual CA5 for 1998 to 2013, and explores their relationships.

Regression analysis is used to answer RQ3. Research by Thelwall and Wilson (2014) showed that a better regression strategy for citation data is to "add one to the citations, take their log and then use the general linear (ordinary least squares) model for regression (e.g., multiple linear regression, ANOVA), or to use the generalized linear model without the log." (p. 963). Thus, this study converts "With or without foreign language references", "Discipline", "Prestige of journal", "Prestige of institution", and "Scientific collaboration" to dummy variables and sets them as independent variables, then sets $log_{10}(C5+1)$ as the dependent variable and uses the multiple linear regression model (see Eq. 3) for analysis. This can be represented as

$$log_{10}(C5_i + 1) = b_0 + b_1 PFLR_i + b_2 Econ_i + b_3 Ed_i + b_4 Evs_i + b_5 Geo_i + b_6 Law_i + b_7 Lis_i \\ + b_8 Mgmt_i + b_9 Pols_i + b_{10} Psy_i + b_{11} Soc_i + b_{12} Sps_i + b_{13} Stat_i \\ + b_{14} LeadingJournal_i + b_{15} TopInstitution_i + b_{16} Coauthorship_i + \varepsilon_i$$

(3)

where PFLR$_i$=1 if the paper i is coded as "Papers with foreign language references" (PFLR), and PFLR$_i$=0 otherwise; Econ$_i$ (Ed$_i$, Evs$_i$, ..., and Stat$_i$) =1 if the paper i is code as "Econ" (Ed, Evs, ..., and Stat), and Econ$_i$ (Ed$_i$, Evs$_i$, ..., and Stat$_i$) =0 otherwise; LeadingJournal$_i$=1 if the paper i is coded as "Leading journals", and LeadingJournal$_i$=0 otherwise; TopInstitution$_i$=1 if the paper i is code as "Top institutions", and TopInstitution$_i$=0 otherwise; Coauthorship$_i$=1 if the paper i is coded as "Co-authorship", and Coauthorship$_i$=0 otherwise.

## 3. Results

### 3.1. Do papers with foreign language references receive more citations?

Table 3 shows the statistical results of the basic indicators. Among the 950,302 Chinese social sciences papers, papers with foreign language references (PFLR) account for 35.5% of all papers. However, total citations received by PFLR in two years (∑C2) account for 57.0%, which is more than those received by papers without foreign language references (PNLR). The gap between PFLR and PNLR widens in the sum of 5-year cumulative citation count (∑C5), with the proportion of citations received by PFLR rising to 62.5%. This demonstrates that although PFLR make up the minority of publications, they receive the majority of citations.

**Table 3** Total number of publications and citations for all papers, papers with foreign language references (PFLR), and papers without foreign language references (PNLR)

|  | Number of publications | Number of citations, two-year citation window (∑C2) | Number of citations, five-year citation window (∑C5) |
| --- | --- | --- | --- |
| All papers | 950,302 | 518,113 | 1,180,043 |
| PFLR | 337,043 | 295,291 | 737,880 |
| (%PFLR) | 35.5 | 57.0 | 62.5 |
| PNLR | 613,259 | 222,822 | 442,163 |
| (%PNLR) | 64.5 | 43.0 | 37.5 |

The 2-year cumulative citations per paper (CPP2) and the 5-year cumulative citations per paper (CPP5) of all papers, PFLR, and PNLR are presented in Figure 2. It can be seen that both CPP2 and CPP5 of PFLR are higher than those of PNLR. In terms of the increase from CPP2 to CPP5, PFLR is also higher than PNLR, which shows that papers with foreign language references not only receive more citations on average, but also accumulate citations faster over the time period studied.

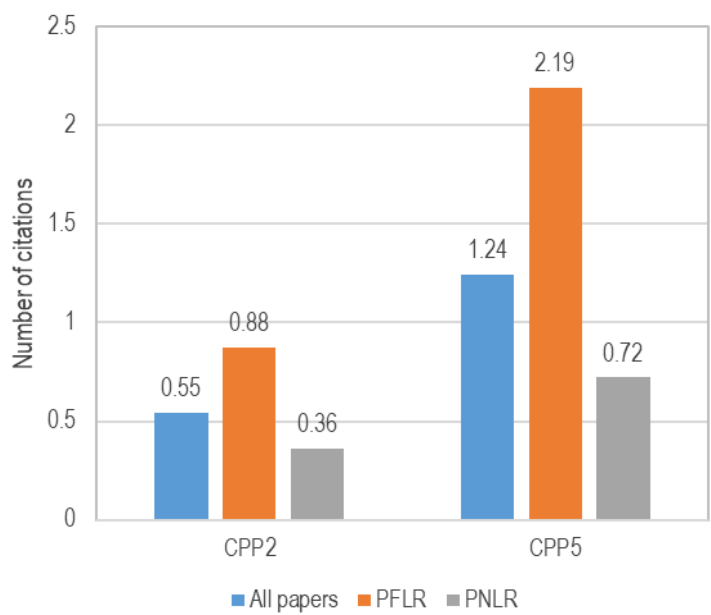

**Fig. 2** Average number of citations (CPP2 and CPP5) for all papers, papers with foreign language references (PFLR), and papers without foreign language references (PNLR)

In order to quantify the impact of foreign language references on citations, the 2-year cumulative citation advantage (CA2) and the 5-year cumulative citation advantage (CA5) are calculated. The results are CA2=2.41 and CA5=3.04, which show that papers with foreign language references are, on average, cited more than twice as much as PNLR. In addition, the result that CA5 is greater than CA2 demonstrates that the gap between average citation count of PFLR and that of PNLR becomes larger over time, suggesting that the former has a more lasting academic impact than the latter. Taking "With or without foreign language references" as the grouping variable, Mann-Whitney U test is separately performed on C2 and C5. As shown in Table 4, in both tests for C2 and C5, there is a significant difference between the citation count of PFLR and that of PNLR ($p<0.001$).

**Table 4** Difference of citation count (C2 and C5) between papers with foreign language references (PFLR) and papers without foreign language references (PNLR), compared using a Mann-Whitney U test [a]

|  | C2 | C5 |
|---|---|---|
| Mann-Whitney U | 8.345E+10 | 7.264E+10 |
| Wilcoxon W | 2.715E+11 | 2.607E+11 |
| Z | -198.822 | -271.637 |
| Asymp. Sig. (2-tailed) | .000 | .000 |

a. Grouping Variable: With or without foreign language references

Based on the above analyses, the answer to RQ1 is that the papers with foreign language references receive more citations than those without, and that this gap is wider in terms of long-term citations.

### 3.2. Does the CA decrease as the %PFLR increases?

Figure 3A presents the annual proportion of papers with foreign language references (%PFLR) for 1998 to 2013. It can be seen that %PFLR continuously increased from 1998 to 2013. In 1998, only 11.7% of Chinese social sciences papers cited foreign language literature. However, the percentage grew to 54.4% in 2013, which means that more than half of these papers published in 2013 have cited foreign language literature.

Figure 3B shows the diachronic trends of the 2-year cumulative citation advantage (CA2) and the 5-year cumulative citation advantage (CA5). It can be seen that both CA2 and CA5 show downward trends before 2008, which indicates that the advantage of PFLR in citations gradually narrowed from 1998 to 2008. In other words, the positive citation impact of citing foreign language literature has gradually weakened from 1998 to 2008. However, after 2008, both CA2 and CA5 were no longer in continuous decline, but remained relatively stable around 2 and 2.5. It demonstrates that there is a continuous relationship between citing foreign literature and receiving citations, and the number of citations received by PFLR is always about twice or more than that received by PNLR.

In response to RQ2, we find, by combining the Figure 3A and B, that the citation advantage of papers with foreign language references decreased as their proportion increased in the early years (from 1998 to 2008). But after 2008, although the proportion of papers with foreign language references continued to increase, their citation advantage still existed and remained relatively stable.

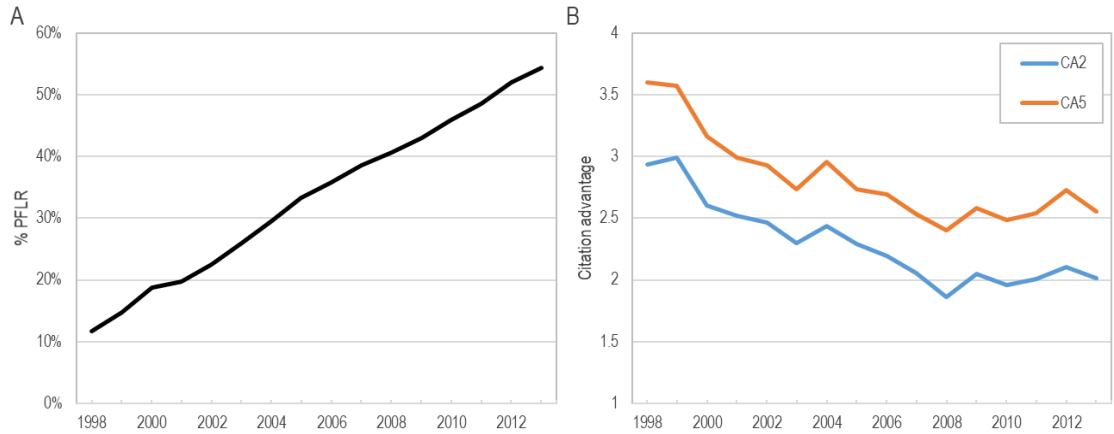

Figure 3. A) Percentage of papers with foreign language references (%PFLR) and B) 2-year (CA2) and 5-year (CA5) cumulative citation advantage from 1998 to 2013

## 3.3. Does the CA still exist after controlling for other characteristics of citing papers?

### 3.3.1 Discipline

The analyses presented above focused on the social sciences as a whole. However, subfields of the social sciences vary in their research objects and citation behaviors (Bornmann and Daniel 2008; Tahamtan et al. 2016). Scholars working on localized objects and topics may naturally focus more on the domestic knowledge source (Gingras and Mosbah-Natanson 2010; Larivière 2018; Warren 2004). On the other hand, the difference of citation behavior also leads to the interdisciplinary gap in the possibility of receiving citations, which is the reason why journal impact factors cannot be compared across disciplines (Dorta-González and Dorta-González 2013; Sugimoto and Larivière 2018). To address these differences, we examine the relationship between foreign language citing and citations received for papers in each of the 13 disciplines.

The proportion of papers citing foreign language references (%PFLR) of different disciplines is shown in Figure 4. Papers in psychology are most likely to draw upon foreign language references, with a %PFLR of 81.6%. This is contrasted with Journalism & Communication, which is at 10.7% -- less than one seventh of the rate of Psychology. About half of the papers from Management (%PFLR=48.0%) cite foreign literature. Considering the large amount of Chinese papers on Management, it suggests higher rates of international engagement with literature for this discipline. In addition to Journalism & Communication, other disciplines with lower %PFLR include Sports Science (%PFLR=30.0%), Political Science (%PFLR=29.2%), and Education (%PFLR=27.2%), indicating Chinese scholars largely rely on local language papers in these knowledge communities. These results converge with those obtained by He (2008): disciplines in which researchers from China are more likely to publish in SSCI- and A&HCI-indexed journals rather than CSSCI-indexed ones are also those with a larger %PFLR.

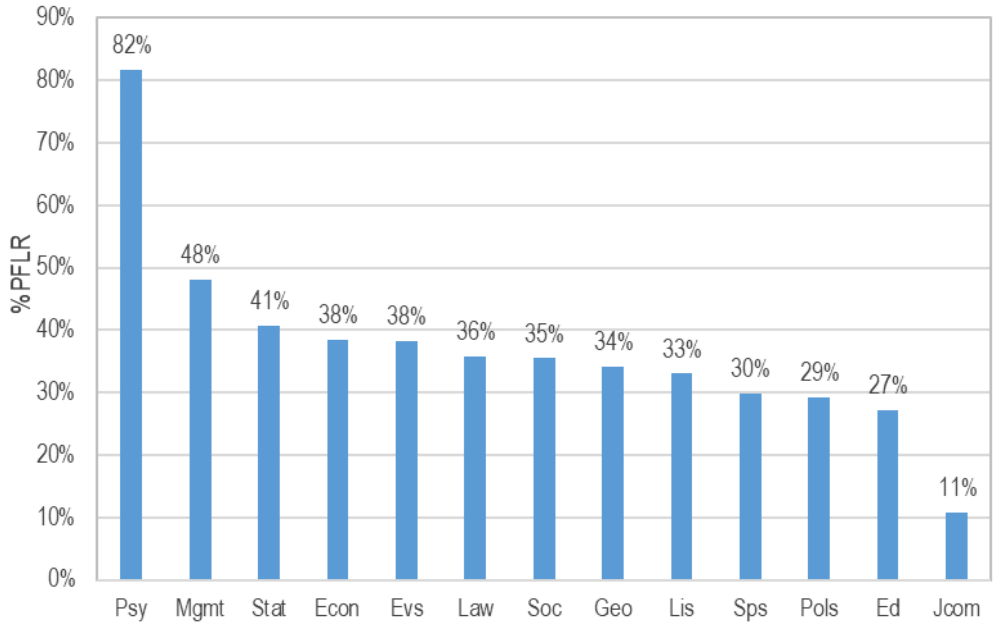

**Fig. 4** Percentage of papers with foreign language references (%PFLR), by discipline

Figure 5 and 6 present the 2-year (CPP2) and 5-year (CPP5) cumulative citations per paper for all papers, papers with foreign language references (PFLR), and those without foreign language references (PNLR) in each discipline. While citation rates vary across disciplines, results show that, in each discipline, papers with foreign language references obtain more citations than those without such references.

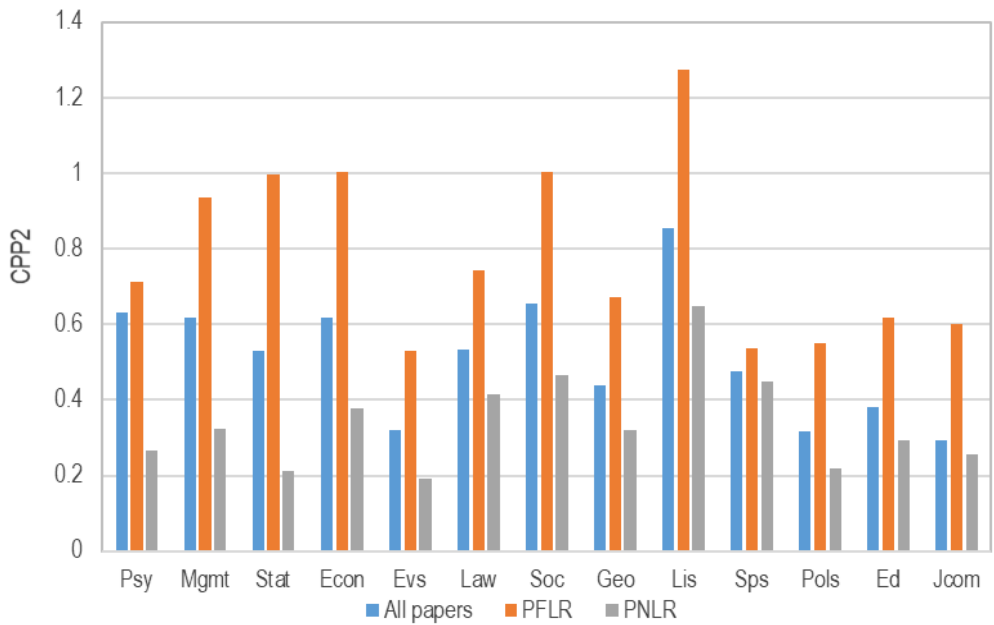

**Fig. 5** 2-year cumulative citations per paper (CPP2) for all papers, papers with foreign language references (PFLR) and papers without foreign language references (PNLR), by discipline

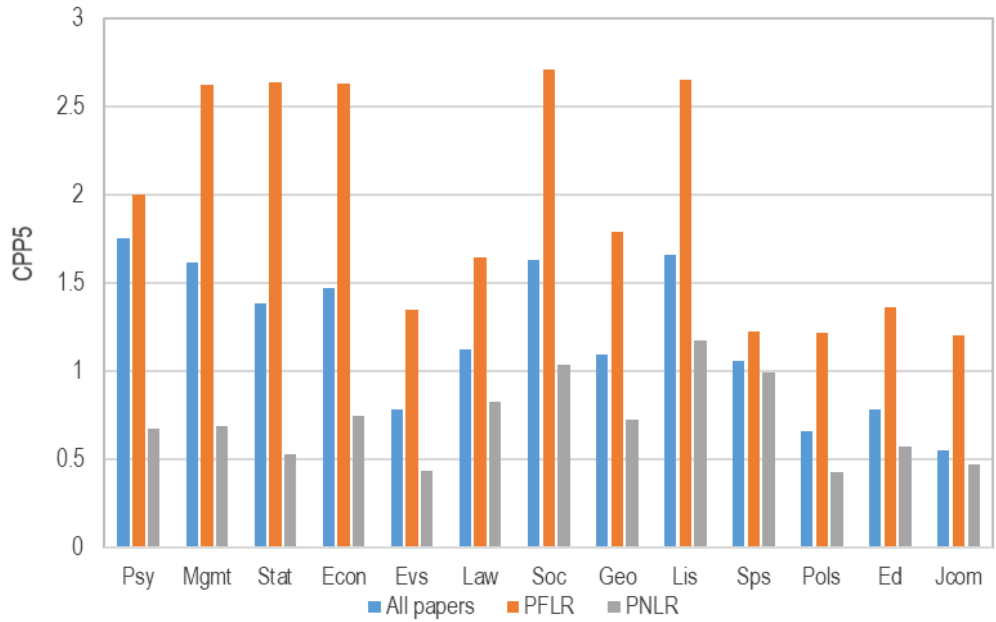

**Fig. 6** 5-year cumulative citations per paper (CPP5) for all papers, papers with foreign language references (PFLR) and papers without foreign language references (PNLR), by discipline

Figures 5 and 6 show that, regardless of in which discipline, the CPP2 of PFLR is always greater than that of PNLR, and so is the CPP5. This demonstrates that, despite disciplinary differences in citation rates, there remains an invariant relationship between citing foreign language literature and receiving citations. In order to further quantify the citation advantage of PFLR in each discipline, the citation advantages (CA2 and CA5) are calculated by discipline (Table 5). It can be seen that the CA2 of all disciplines except Law, Library & Information Science, and Sports Science are greater than two, which means that the average citation count of PFLR is more than twice larger than that of PNLR. Sports Science has the lowest CA2 at 1.20, but still greater than 1. Disciplines with larger CA2 also have the larger CA5, and the CA5 of Sports Science is still the smallest at 1.24. In addition, the citation advantage of PFLR expands with the extension of citation window in all disciplines, which suggests that PFLR have a faster citation speed in all disciplines. Taking "With or without foreign language references" as the grouping variable, Mann-Whitney U test is separately performed on the C2 and C5 in each discipline. As shown in Table 6, in all tests for C2 and C5, there is a significant difference between the citation count of PFLR and PNLR ($p<0.001$). On the whole, in all disciplines, papers with foreign language references receive more citation than those without.

**Table 5** 2-year (CA2) and 5-year (CA5) cumulative citation advantage, by discipline

| Discipline | Psy | Mgmt | Stat | Econ | Evs | Law | Soc | Geo | Lis | Sps | Pols | Ed | Jcom |
|---|---|---|---|---|---|---|---|---|---|---|---|---|---|
| CA2 | 2.67 | 2.89 | 4.68 | 2.67 | 2.77 | 1.79 | 2.16 | 2.11 | 1.97 | 1.20 | 2.51 | 2.11 | 2.35 |
| CA5 | 2.97 | 3.81 | 5.00 | 3.52 | 3.09 | 1.99 | 2.60 | 2.46 | 2.25 | 1.24 | 2.83 | 2.39 | 2.56 |

**Table 6** Difference of citation count (C2 and C5) between papers with foreign language references (PFLR) and papers without foreign language references (PNLR) in each discipline, compared using a Mann-Whitney U test [a]

| Discipline | C2 | | | | C5 | | | |
|---|---|---|---|---|---|---|---|---|
| | Mann-Whitney U | Wilcoxon W | Z | Asymp. Sig. (2-tailed) | Mann-Whitney U | Wilcoxon W | Z | Asymp. Sig. (2-tailed) |
| Psy | 2.000E+07 | 2.570E+07 | -22.643 | .000 | 1.677E+07 | 2.247E+07 | -32.298 | .000 |
| Mgmt | 1.433E+09 | 3.450E+09 | -86.556 | .000 | 1.196E+09 | 3.213E+09 | -119.11 | .000 |
| Stat | 2.274E+06 | 7.056E+06 | -24.98 | .000 | 1.807E+06 | 6.589E+06 | -31.575 | .000 |
| Econ | 9.398E+09 | 2.856E+10 | -128.567 | .000 | 8.043E+09 | 2.721E+10 | -174.844 | .000 |
| Evs | 4.452E+07 | 1.294E+08 | -27.539 | .000 | 4.019E+07 | 1.250E+08 | -35.539 | .000 |
| Law | 5.626E+08 | 1.730E+09 | -37.713 | .000 | 5.185E+08 | 1.686E+09 | -50.881 | .000 |
| Soc | 1.389E+08 | 4.489E+08 | -37.156 | .000 | 1.232E+08 | 4.332E+08 | -49.498 | .000 |
| Geo | 2.781E+06 | 9.141E+06 | -12.047 | .000 | 2.399E+06 | 8.759E+06 | -18.119 | .000 |
| Lis | 6.009E+08 | 2.121E+09 | -53.248 | .000 | 5.384E+08 | 2.058E+09 | -70.199 | .000 |
| Sps | 9.052E+07 | 3.143E+08 | -8.992 | .000 | 8.715E+07 | 3.109E+08 | -13.128 | .000 |
| Pols | 5.811E+08 | 2.268E+09 | -54.703 | .000 | 5.187E+08 | 2.206E+09 | -71.978 | .000 |
| Ed | 1.480E+09 | 6.127E+09 | -57.337 | .000 | 1.320E+09 | 5.967E+09 | -80.256 | .000 |
| Jcom | 2.417E+08 | 2.591E+09 | -33.645 | .000 | 2.236E+08 | 2.573E+09 | -42.251 | .000 |

a. Grouping Variable: With or without foreign language references

### 3.3.2 Prestige of journal

The proportion of papers with foreign language references (%PFLR) in leading journals is 44.8%, and the %PFLR of other journals is 35.0%. Figures 7A and B show the 2-year cumulative citations per paper (CPP2) and the 5-year cumulative citations per paper (CPP5) for all papers, papers with foreign language references (PFLR), and papers without foreign language references (PNLR) by prestige of journal. As for all papers, it can be found that both CPP2 and CPP5 in leading journals are higher than those in other journals, showing that papers published in the leading journals receive more citations than those published in other journals, supporting previous studies (e.g., Callaham et al. 2002; Xie et al. 2018).

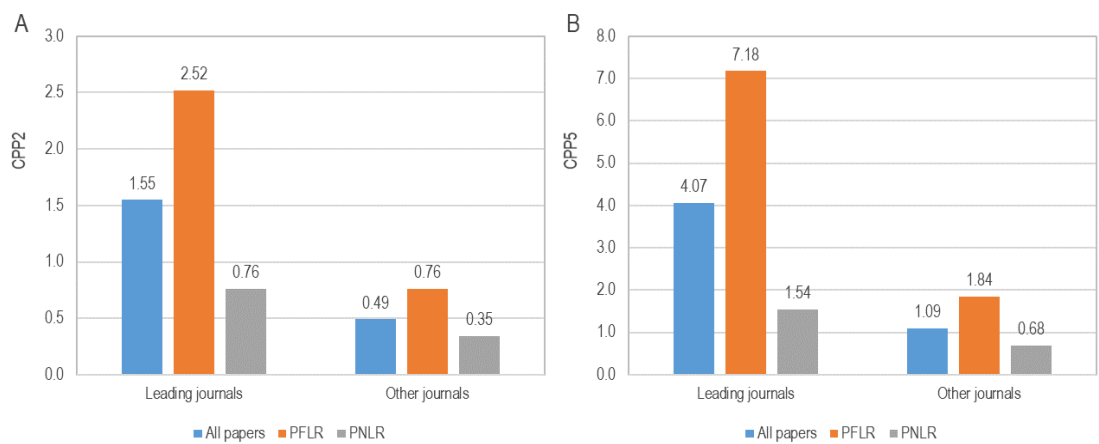

Figure 7. A) 2-year cumulative citations per paper (CPP2) and B) 5-year cumulative citations per paper (CPP5) for all papers, papers with foreign language references (PFLR) and papers without foreign language references (PNLR), by type of journals

As shown in Figure 7, CPP indicators of PFLR are always greater than those of PNLR, irrespective of the citation window. It means that papers with foreign language references always receive more citations than those without in both leading and regular journals. In order to further quantify the advantage, the citation advantages (CA2 and CA5) are calculated. The CA2 of leading journals is 3.32 while the CA2 of other journals is 2.21. The CA5 of leading journals is 4.65, while that of other journals is 2.70. All of the CA are greater than 1, and both CA2 and CA5 of leading journals are much greater than those of other journals, which suggests that the language diversity of references has a greater impact on the citation count of papers in prestigious journals. Taking "With or without foreign language references" as the grouping variable, a Mann-Whitney U test is separately performed on the citation count (C2 and C5) of papers in both kinds of journals. As shown in Table 7, there is a significant difference between the citation count of PFLR and PNLR ($p<0.001$), regardless of the citation window or the prestige of journal.

**Table 7** Difference of citation count (C2 and C5) between papers with foreign language references (PFLR) and papers without foreign language references (PNLR) in leading and other journals, compared using a Mann-Whitney U test [a]

|  | Leading journals | | Other journals | |
| --- | --- | --- | --- | --- |
|  | C2 | C5 | C2 | C5 |
| Mann-Whitney U | 1.876E+08 | 1.499E+08 | 7.589E+10 | 6.635E+10 |
| Wilcoxon W | 5.489E+08 | 5.113E+08 | 2.478E+11 | 2.383E+11 |
| Z | -75.802 | -97.065 | -181.534 | -251.996 |
| Asymp. Sig. (2-tailed) | .000 | .000 | .000 | .000 |

a. Grouping Variable: With or without foreign language references

The above shows that the citation advantage of papers with foreign language references remains after testing for the prestige of journal.

### 3.3.3 Prestige of institution

The proportion of papers with foreign language references (%PFLR) of top institutions is 52.2%, while that of other institutions is 26.7%. The former is higher than the latter, suggesting that papers from top institutions are more likely to cite foreign language references. Figures 8 shows the 2-year cumulative citations per paper (CPP2) and the 5-year cumulative citations per paper (CPP5) for all papers, papers with foreign language references (PFLR), and papers without foreign language references (PNLR), by prestige of institution. As expected, the CPP for top institutions is higher than that of lower-ranked institutions, regardless of the length of citation window. This result is consistent with what previous studies found (e.g., Amara et al. 2015; Mou et al. 2018).

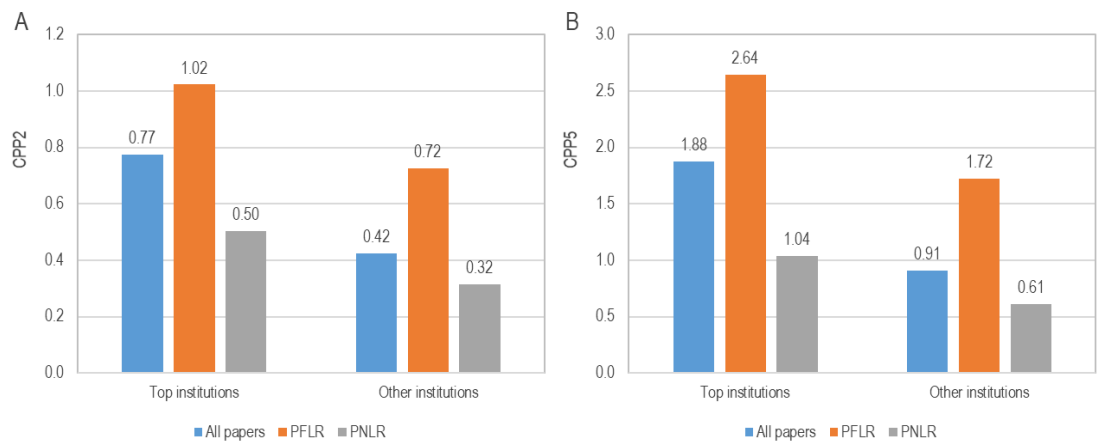

Figure 8. A) 2-year cumulative citations per paper (CPP2) and B) 5-year cumulative citations per paper (CPP5) for all papers, papers with foreign language references (PFLR) and papers without foreign language references (PNLR), by type of institutions

Furthermore, Figure 8 demonstrates that, regardless of institution type, the CPP of PFLR is greater than that of PNLR for both citation windows. This shows that papers with foreign language references, regardless of the prestige of authors' institutions, receive more citations than those without. In order to further quantify the advantage of PFLR in average citation count, the citation advantages are calculated. The CA2 of top institutions is 2.03 and the CA2 of other institution is also 2.30. The CA5 of top institutions is 2.55, while that of other institutions is 2.81. These values show that the average citation count of PFLR is about twice that of PNLR, and the degree of PFLR's advantage is not significantly different between two kinds of institutions. Taking "With or without foreign language references" as the grouping variable, Mann-Whitney U test is separately performed on the citation counts (C2 and C5) of papers published by top institutions and other institutions (Table 8). It shows that, regardless of the prestige of the institution, there is a significant difference between the citation count of PFLR and PNLR ($p<0.001$).

**Table 8** Difference of citation count (C2 and C5) between papers with foreign language references (PFLR) and papers without foreign language references (PNLR) in top and other institutions, compared using a Mann-Whitney U test [a]

|  | Top institutions | | Other institutions | |
|---|---|---|---|---|
|  | C2 | C5 | C2 | C5 |
| Mann-Whitney U | 1.114E+10 | 9.860E+09 | 3.122E+10 | 2.750E+10 |
| Wilcoxon W | 2.342E+10 | 2.214E+10 | 1.354E+11 | 1.317E+11 |
| Z | -99.021 | -140.436 | -143.659 | -195.457 |
| Asymp. Sig. (2-tailed) | .000 | .000 | .000 | .000 |

a. Grouping Variable: With or without foreign language references

From the above, it is shown that the citation advantage of papers with foreign language references still exists after testing for the prestige of institution.

### 3.3.4 Scientific collaboration

The proportion of papers with foreign language references (%PFLR) of co-authorship is 49.4%, while that of single authorship is 26.6%. Compared to sole-authored papers, there are more co-authored papers citing foreign language literature. The 2-year cumulative citations per paper (CPP2) and the 5-year cumulative citations per paper (CPP5) of total papers, papers with foreign language references (PFLR), and papers without foreign language references (PNLR) written by collaboratively and singly are shown in Figure 9. Reinforcing several studies (e.g., Larivière et al. 2015), we show that the CPP in papers written by multi-authors are greater than those by sole author in both 2-year and 5-year windows, indicating that co-authored papers receive more citations than those written by sole authors.

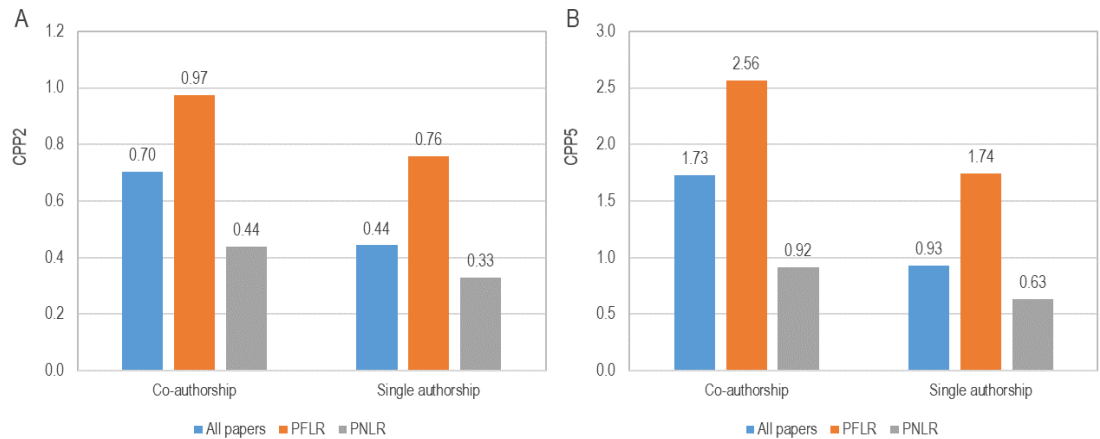

Figure 9. A) 2-year cumulative citations per paper (CPP2) and B) 5-year cumulative citations per paper (CPP5) for all papers, papers with foreign language references (PFLR) and papers without foreign language references (PNLR), by type of authorship

Despite these differences, the CPP of PFLR are always greater than those of PNLR, regardless of the authorship type. The CA2 of co-authorship is 2.21 and that of single authorship is 2.30. The CA5 of co-authorship is 2.80 and that of single authorship is 2.75. The values of citation advantage are quite similar between two kinds of authorship, and the PFLR receive more than twice citations as PNLR. Taking "With or without foreign language references" as the grouping variable, the Mann-Whitney U test is separately performed on the C2 and C5 of papers written by co-authors and sole author, the results are shown in Table 9. It can be found that there is a significant difference between the citation count of papers with foreign language references and those without regardless of the authorship ($p<0.001$).

**Table 9** Difference of citation count (C2 and C5) between papers with foreign language references (PFLR) and papers without foreign language references (PNLR) in co-authored and single-authored papers, compared using a Mann-Whitney U test [a]

|  | Co-authorship | | Single authorship | |
| --- | --- | --- | --- | --- |
|  | C2 | C5 | C2 | C5 |
| Mann-Whitney U | 1.403E+10 | 1.226E+10 | 2.709E+10 | 2.408E+10 |
| Wilcoxon W | 3.169E+10 | 2.992E+10 | 1.175E+11 | 1.145E+11 |
| Z | -117.106 | -163.983 | -135.18 | -181.57 |
| Asymp. Sig. (2-tailed) | .000 | .000 | .000 | .000 |

a. Grouping Variable: With or without foreign language references

In conclusion, after testing for scientific collaboration, papers with foreign language references still receive more citations than those without.

### 3.3.5 Regression analysis

Various factors analyzed above were also combined in a multiple liner regression analysis to investigate their relationship to citations. Table 10 shows that the regression model is statistically significant (F(16, 950285)= 9320.659, p<0.001) and approximately 13.6% of variance in citations can be explained by with or without foreign language references, discipline, prestige of journal, prestige of institution, and scientific collaboration ($R^2$=0.136). It is obvious that the widely-recognized characteristics still have significant influence on citations in this regression, specifically, there are significant differences in the number of citations received by papers in different disciplines (p<0.001), papers published in leading journals receive more citations than those published in other journals (β=0.144, p<0.001), papers from top institutions have more citations than those from other institutions (β=0.097, p<0.001), and papers written by co-authors receive more citations than those written by sole author (β=0.085, p<0.001). The most important finding in this regression is that after controlling all of above characteristics, paper with foreign language references (PFLR) still have more citations than papers without foreign language references (PNLR) (β=0.229, p<0.001). In addition, the β of PFLR is the largest among all variables, which suggests that compared with the difference in discipline, in prestige of journal and institution, as well as in scientific collaboration, citing foreign language literature has a stronger influence on a paper's citations.

To answer to RQ3, it is clear that the citation advantage of papers with foreign language references remains after controlling the widely-recognized characteristics of citing papers.

Table 10 Result of the multiple liner regression analysis

| Independent Variable | Standardized Coefficients (β) |
|---|---|
| With or without foreign language references | |
| *PFLR* | .229*** |
| Discipline | |
| *Econ* | .057*** |
| *Ed* | -.007*** |
| *Evs* | -.027*** |
| *Geo* | .008*** |
| *Law* | .026*** |
| *Lis* | .093*** |
| *Mgmt* | .025*** |
| *Pols* | -.041*** |
| *Psy* | .004*** |
| *Soc* | .042*** |
| *Sps* | .021*** |
| *Stat* | .006*** |
| Prestige of journal | |
| *LeadingJournal* | .144*** |
| Prestige of institution | |
| *TopInstitution* | .097*** |
| Scientific collaboration | |
| *Coauthorship* | .085*** |
| $R^2$=.136    F(16, 950285)= 9320.659    Sig.=.000 | |

*** p<0.001, N=950,302

# 4. Discussion and Conclusion

Scientometricians have long relied on a few citation databases to inform their understanding of the progress of science. However, these databases have widely observed biases in favor of English-speaking publications and journals from western countries. The present research provides a novel contribution by taking a database focused on Chinese authors and publications. This nationally-oriented database provides a rare analysis of the consumption of foreign literature and the impact of the resulting work on the domestic research community.

We demonstrate a stark increase in the consumption of foreign language material within the Chinese social science community: in 1998, only 11.7% of papers cited foreign language literature; this number increased to more than half in 2013. Of all the foreign language references, English accounts for the vast majority (96.4%). Citations are often noted for their symbolic function, but they also serve a strong instrumental function in signaling to readers' works with which they may be unaware (Merton, 1988). This is particularly the case for foreign-language material: citing relevant work may serve to bridge scientific communities across countries. In this case, they become a window for domestic scholars to understand the work of international peers and thus play a key media role in the diffusion of knowledge from the international to the local.

There is a significant citation advantage for papers that play this role: those papers incorporating foreign language literature receive significantly more citations than those that do not. The citation advantage of internationalized work holds when controlling for characteristics of the citing papers, such as the discipline, prestige of journal, prestige of institution, and scientific collaboration. Given that the citations are calculated internally—that is, citations coming from the Chinese social science community—this suggests a higher scientific impact of papers that are absorbing foreign literature. Therefore, limiting access to foreign language material within China will have significant implications in the production of high-impact work.

As we observed, the citation advantage has decreased from 1998 to 2008, largely as an artifact of the increased number of papers citing foreign language material. This may demonstrate the homogenizing effect of globalization: as papers draw upon the same sources of knowledge, the disparities in impact will lessen. However, this move towards globalization can also raise several issues, particularly when it is oriented towards a single language. Adopting a lingua franca in science can allow for easier communication across cultures, but it can also have a chilling effect on local research topics which are not valued by the cultures of the dominant language. However, we can also observe that after 2008, although the proportion of papers with foreign language references continued to increase, the citation advantage no longer declines and is now relatively stable. This result suggests that incorporating foreign language literature continues to increase scientific impact, even as the scientific community itself becomes increasingly international. It is important to note that innovation does not only occur in a single language. For example, papers that earned their authors a Nobel Prize have been written in 25 different languages over the last century (1901-2017) (Nobelprize.org 2018). Only a quarter of the prize-winning papers were in English, the lingua franca of science for most of that period (Gordin 2015). Therefore, it is critical to remain mindful of the diversity of languages that can allow for the broadest diversity of investigation in science.

# Acknowledgments

This research was financially supported by the National Social Science Fund of China (No. 17BTQ014) and the China Scholarship Council. We'd like to thank Fansai Meng for collecting data and Yi Bu as well as Jing Yang for fruitful discussions and comments. We also appreciate two anonymous reviewers for their helpful suggestions.